\documentclass[prd,twocolumn,showpacs,preprintnumbers,amsmath,amssymb]{revtex4}

\usepackage{graphics}
\usepackage{epsfig}
\usepackage{dcolumn}
\usepackage{bm}

\begin{document}

\title{Is $Z^+(4430)$ a loosely bound molecular state?}

\author{Xiang Liu$^1$}
\email{xiangliu@pku.edu.cn}

\author{Yan-Rui Liu$^{2}$}
\email{yrliu@ihep.ac.cn}

\author{Wei-Zhen Deng$^1$}
\email{dwz@th.phy.pku.edu.cn}
\author{Shi-Lin Zhu$^1$}
\email{zhusl@phy.pku.edu.cn}

\affiliation{$^1$Department of Physics, Peking University, Beijing
100871, China \\ $^2$Institute of High Energy Physics, P.O. Box
918-4, Beijing 100049, China}

\date{\today}

\begin{abstract}

Since $Z^+(4430)$ lies very close to the threshold of $D^\ast{\bar
D}_1$, we investigate whether $Z^+(4430)$ could be a loosely bound
S-wave state of $D^\ast{\bar D}_1$ or $D^\ast{\bar D}^\prime_1$
with $J^P=0^-, 1^-, 2^-$, i.e., a molecular state arising from the
one-pion-exchange potential. The potential from the crossed
diagram is much larger than that from the diagonal scattering
diagram. With various trial wave functions, we notice that the
attraction from the one pion exchange potential alone is not
strong enough to form a bound state with realistic pionic coupling
constants deduced from the decay widths of $D_1$ and $D^\prime_1$.

\end{abstract}


\maketitle

\section{introduction}\label{sec1}

Recently Belle Collaboration observed a sharp peak in the
$\pi^{+}\psi'$ invariant mass spectrum in the exclusive $B\to K
\pi^{+}\psi'$ decays with a statistical significance of $7\sigma$
\cite{Belle-4430}. This resonance-like structure is named as
$Z^{+}(4430)$. The fit with a Breit-Wigner form yields its mass
$m=4433\pm4(\mathrm{stat})\pm1(\mathrm{syst})$ MeV and a narrow
width
$\Gamma=44^{+17}_{-13}(\mathrm{stat})^{+30}_{-11}(\mathrm{syst})$
MeV. It is very interesting to note that the width of $Z^{+}(4430)$ is
roughly the same as that of $D_1$.

The product branching fraction is measured to be ${\cal
B}(B\to KZ^+(4430))\cdot {\cal B}(Z^+(4430)\to\pi^+\psi')=(4.1\pm
1.0(\mathrm{stat})\pm1.3(\mathrm{syst}))\times 10^{-5}$ \cite{Belle-4430}.
For comparison, we list the production rate of $X(3872)$ and $Y(4260)$
in $B$ decays. From Ref. \cite{Barbar-BR3872} we have
\begin{eqnarray*}
&&{\cal B}(B^-\rightarrow K^-X(3872))\cdot {\cal
B}(X(3872)\rightarrow J/ \Psi\pi^+\pi^-)\nonumber\\
&=&(1.28\pm 0.41)\times 10^{-5},
\end{eqnarray*}
and from Ref. \cite{Barbar-BR3872-06}
\begin{eqnarray*}
&&{\cal B}(B^-\rightarrow K^- X(3872))\cdot {\cal B}(X(3872)
\rightarrow J/ \Psi
\pi^+\pi^-)\nonumber\\
&=&(10.1\pm 2.5\pm 1.0)\times 10^{-6}.
\end{eqnarray*}
For $Y(4260)$, Babar Collaboration gave the upper limit of the
branching fraction \cite{Barbar-BR3872-06}
\begin{eqnarray*}
&&{\cal B}(B^-\rightarrow K^- Y(4260))\cdot {\cal
B}(Y(4260)\rightarrow J/ \Psi \pi^+\pi^-)\nonumber\\
&<&2.9\times 10^{-5}.
\end{eqnarray*}
It's plausible that (1) ${\cal B}(B\to KZ^+(4430))$ is comparable to
both ${\cal B}(B^-\rightarrow K^-X(3872))$ and
${\cal B}(B^-\rightarrow K^- Y(4260))$; (2) $\pi^+\psi^\prime$ is one
of the main decay modes of $Z^+(4430)$ if it is a resonance.

The peak $Z^{+}(4430)$ inspired several theoretical speculations
of its underlying structure. Rosner suggested that $Z^+(4430)$ is
a S-wave threshold effect because $Z^{+}(4430)$ lies close to the
$D^*(2010)\bar{D}_{1}(2420)$ threshold \cite{rosner}. The
production mechanism was speculated as follows. The $b$ quark
first decays into a strange quark and a pair of $c\bar c$ while a pair
of light quarks are created from the vacuum. In other words, $B$
meson first decays into a $K$ and a pair of $D$ mesons. Then the
$D$ meson pair re-scatters into $\pi^{+}\psi'$. He also suggested
other possible charged states.

Maiani, Polosa and Riquer identified this signal as the first
radial excitation of the tetraquark supermultiplet to which
$X(3872)$ and $X(3876)$ belong \cite{maiani}. With their
assignment the quantum number of $Z^{+}(4430)$ is $J^{PC}=1^{+-}$
where the C-parity is for the neutral member within the same
multiplet. $Z^{+}(4430)$ decays into $\pi^{+}\psi'$ via S-wave.
Moreover their scheme requires the ground state with
$J^{PC}=1^{+-}$ around 3880 MeV which decays into $\pi^{+}\psi$
and $\eta_c \rho^+$.

With a QCD-string model, Gershtein, Likhoded and Pronko argued
that both X(3872) and $Z^+(4430)$ are tetraquark states
\cite{Gershtein}. They speculate that the two quarks and two
anti-quarks sit on the four corners of a
square while any $q\bar q$ pair is a color-octet
state. The decays of tetraquarks involve the reconnection
of the color string.

Cheung, Keung and Yuan discussed the bottom analog of $Z^+(4430)$
assuming it is a tetraquark bound state \cite{cky}. According to
their estimate, the doubly-charged $Z_{bc}$ state lies
around 7.6 GeV while the bottomonium analog $Z_{bb}$ of $Z^+
(4433)$ is about 10.7 GeV.

Qiao suggested $Z^+(4433)$ be the first radial excitation of
$\Lambda_c -\Sigma_c^0$ bound state \cite{Qiao}. Within Qiao's
scheme, all of the recently observed states $Y(4260)$, $Y(4361)$,
$Z^+(4430)$ and $Y(4664)$ are accommodated in the extended heavy
baryonium framework.

Lee, Mihara, Navarra and Nielsen calculated the mass of
$Z^+(4430)$ $m_Z=(4.40\pm0.10)$ GeV in the framework of QCD sum
rules, assuming it is a $0^-$ molecular state of
$D^*(2010)\bar{D}_{1}(2420)$ \cite{qsr-lee}. They predicted the
analogous mesons $Z_{s}$ at $m_{Z_{s}}= (4.70\pm 0.06)$ GeV, which
is above the $D_s^*D_1$ threshold and $Z_{bb}$ around $m_{Z_{bb}}=
(10.74\pm 0.12)$ GeV.

Bugg proposed that $Z^+(4430)$ is a threshold cusp arising from
the deexcitation of the $D^*(2010)D_1(2420)$ pair into lower mass
D states \cite{Bugg}. The imaginary part $Im f(s)$ of the elastic
S-wave amplitude $f(s)$ is a step function near the threshold.
From the dispersion relation, the real part of the amplitude near
the threshold looks like $Re f(s) \sim \int_{s_0} {\theta (s)\over
s'-s}ds'\sim \ln (s-s_0)$. Therefore $|f(s)|^2$ contains a sharp
cusp near threshold.

However, none of the above schemes explains why $Z^{+}(4430)$ does
{\sl NOT} decay into $\pi^{+}\psi$. One notes that the momentum of
$D$ and ${\bar D}$ is small and close to each other in the rest
frame of the parent $B$ meson, especially when one (or two) of the
$D$ meson pair is an excited state. Therefore, there is plenty of
time for the $D$ meson pair to move together and re-scatter into
$\pi^{+}\psi'$. However, the most puzzling issue of the
re-scattering mechanism is the absence of any signal in the
$\pi^{+}J/\psi$ channel. One may wonder whether the mismatch of
the Q-values of the initial and final states plays an important
role. If so, one should also expect a signal in the $\pi^{+}\psi
(3S)$ channel since there is nearly no mismatch of the Q-value
now. Another potential scapegoat is the specific nodal structure
in the wave functions of the final states. Detailed calculations
along the above two directions are highly desirable to investigate
the origin of the non-observation of $Z^{+}(4430)$ in the
$\pi^{+}\psi$ mode.

In the heavy quark limit, the angular momentum of the light quark
$j_l={\vec l}+{\vec S}_q$ is a good quantum number where $l$ is
the orbital angular momentum and $S_q$ is the light quark spin.
For the P-wave heavy mesons, $j_l={3\over 2}$ or ${1\over 2}$,
which correspond to the two doublets with $J^P=(1^+, 2^+)$ and
$(0^+, 1^+)$ respectively. The ground states D and $D^\ast$ belong
to the $J^P=(0^-, 1^-)$ doublet with $j_l={1\over 2}$. Since the
$1^+$ state in the $(1^+, 2^+)$ doublet decays into $D^\ast \pi$
via D-wave, it's very narrow and denoted as $D_1(2420)$
\cite{PDG}. The $1^+$ state in the $(0^+, 1^+)$ doublet decays
into $D^\ast \pi$ via S-wave. Hence it's very broad and denoted as
$D_1'(2430)$ \cite{PDG}.

Assuming $Z^+(4430)$ is a $D_1D^*$ (or $D_1'D^*$) S-wave
resonance, Meng and Chao found that the open-charm decay mode
$D^*D^*\pi$ is dominant and the re-scattering effects are
significant in $D_1D^*$ channel but not in $D_1'D^*$ channel since
$D_1'$ is very broad \cite{Meng}. For the $J^P=1^-$ candidate, the
ratio $\Gamma(Z^+\to\psi'\pi^+)/\Gamma(Z^+\to J/\psi\pi^+)$ may
reach $5.3$ with a special set of parameters, which partly
accounts for why the $Z^+(4430)$ is difficult to be found in
$J/\psi\pi^+$.

Despite so many theoretical speculations proposed above, a
dynamical study of the $Z^+(4430)$ signal is still missing. In
this work we will explore whether $Z^+(4430)$ could be a S-wave
molecular state of $D^*$ and $\bar{D}_1'$ (or $\bar{D}_1$), which
is loosely bound by the long-range pion exchange potential. We
want to find out whether there exists the attractive
force between $D^*$ and $\bar{D}_1'$ ($\bar{D}_1$) in different
channels.

This paper is organized as follows. We discuss the possible quantum
numbers of $Z^+(4430)$ and its possible partner states in Section
\ref{sec2}. We collect the effective Lagrangians and various
coupling constants in Section \ref{sec3}. We derive the
one-pion-exchange potential (OPEP) in section \ref{sec4}. Then we
present our numerical result and a short discussion in Section
\ref{sec5}.

\section{Quantum number of $Z^+(4430)$ and other possible
states}\label{sec2}

Natively, the smaller its angular momentum, the lower the mass of
$Z^+(4430)$. Since $Z^+(4430)$ lies very close to the
$D^*(2010)\bar{D}_{1}(2420)$ threshold, we consider only the
possibility of $Z^+(4430)$ being the loosely bound S-wave state of
$D^*$ and $\bar{D}_1'$ (or $\bar{D}_1$). Therefore, its possible
angular momentum and parity are $J^P=0^-, 1^-, 2^-$. Moreover,
$Z^{+}(4430)$ was observed in the $\psi'\pi^{+}$ channel. So it is
an isovector state with positive $G$-parity, i.e., $I^G=1^+$.

For a charged member within a molecular isovector multiplet, one
can construct its flavor wave function with definite $G$ parity in
the following way. Suppose $|A\rangle$ is one component of its
flavor wave function. Then the $G=+$ state reads:
\begin{equation}
|+\rangle ={1\over \sqrt{2}}\left(|A\rangle +{\hat G}|A\rangle
\right)
\end{equation}
where ${\hat G}=e^{iI_y\pi}{\hat C}$ is the $G$-parity operator.
Similarly, the $G=-$ state reads:
\begin{equation}
|-\rangle ={1\over \sqrt{2}}\left(|A\rangle -{\hat G}|A\rangle
\right)
\end{equation}

In the present case, the flavor wave function of $Z^{+}(4430)$ is
\begin{eqnarray}
|Z^{+}\rangle=\frac{1}{\sqrt{2}}\big[|A'\rangle+|B'\rangle\big]
\end{eqnarray}
with $|A'\rangle=|\bar{D}_{1}'^{0}D^{*+}\rangle$ and
$|B'\rangle=|D_{1}'^{+}\bar{D}^{*0}\rangle$, where ${D}_{1}'^{0}$
and $D_{1}'^{+}$ belong to the $(0^+, 1^+)$ doublet in the heavy
quark effective field theory. Or
\begin{eqnarray}
|Z^{+}\rangle=\frac{1}{\sqrt{2}}\big[|A\rangle+|B\rangle\big]
\end{eqnarray}
with $|A\rangle=|\bar{D}_{1}^{0}D^{*+}\rangle$ and
$|B\rangle=|D_{1}^{+}\bar{D}^{*0}\rangle$, where ${D}_{1}^{0}$ and
$D_{1}^{+}$ belong to the $(1^+, 2^+)$ doublet.

\begin{figure}[htb]
\scalebox{0.4}{\includegraphics{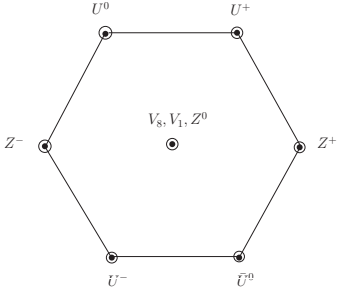}}\caption{The multiplets
composed with charmed mesons and anticharmed mesons. \label{eigen}}
\end{figure}

The flavor wave functions of the partner states of $Z^+$ can be
derived in the following way. Charmed mesons belong to the
fundamental representation of flavor $SU(3)$. Therefore, the
system with a charmed meson and an anti-charmed meson belongs to
$\mathbf{3}\times \mathbf{\bar{3}}=\mathbf{8}+ \mathbf{1}$. We
list the wave functions of these hidden charm states, whose names
are listed in Fig. \ref{eigen}. Here we use the system of
$D_{(s)}^*$ and $D_{(s)1}$ as an illustration.

The flavor wave functions of these states are
\begin{eqnarray*}
|Z^{+}\rangle&=&\frac{1}{\sqrt{2}}\Big(|\bar{D}_{1}^0
D^{*+}\rangle-c|\bar{D}^{*0}D_{1}^+\rangle\Big),\\
\end{eqnarray*}
\begin{eqnarray*}
|Z^{0}\rangle&=&\frac{1}{{2}}\Big[\Big(|{D}_{1}^-
D^{*+}\rangle-c|{D}^{*-}D_{1}^+\rangle\Big)\\
&&-\Big(|\bar{D}_{1}^{0}D^{*0}\rangle-c|\bar{D}^{*0}D_{1}^0\rangle\Big)\Big],\\
|Z^{-}\rangle&=&-\frac{1}{\sqrt{2}}\Big(|{D}_{1}^-
D^{*0}\rangle-c|{D}^{*-}D_{1}^0\rangle\Big),\\
|U^{+}\rangle&=&-\frac{1}{\sqrt{2}}\Big(|\bar{D}_{1}^0
D_s^{*+}-c|\bar{D}^{*0}D_{s1}^+\rangle\rangle\Big),\\
|U^{0}\rangle&=&-\frac{1}{\sqrt{2}}\Big(|{D}_{1}^-
D_s^{*+}\rangle-c|{D}^{*-}D_{s1}^+\rangle\Big),\\
|U^{-}\rangle&=&-\frac{1}{\sqrt{2}}\Big(|{D}_{s1}^-
D^{*0}\rangle-c|{D}_s^{*-}D_{1}^0\rangle\Big),\\
|\bar{U}^{0}\rangle&=&\frac{1}{\sqrt{2}}\Big(|{D}_{s1}^-
D^{*+}\rangle-c|{D}_s^{*-}D_{1}^+\rangle\Big),
\end{eqnarray*}
\begin{eqnarray*}
|V_8\rangle&=&\frac{1}{{2\sqrt3}}\Big[\Big(|{D}_{1}^-
D^{*+}\rangle-c|{D}^{*-}D_{1}^+\rangle\Big)\\
&&+\Big(|\bar{D}_{1}^{0}D^{*0}\rangle-c|\bar{D}^{*0}D_{1}^0\rangle\Big)\\
&&-2\Big(|{D}_{s1}^-
D_s^{*+}\rangle-c|D_s^{*-}D_{s1}^+\rangle\Big)\Big],\\
|V_1\rangle&=&\frac{1}{{\sqrt6}}\Big[\Big(|{D}_{1}^-
D^{*+}\rangle-c|{D}^{*-}D_{1}^+\rangle\Big)\\
&&+\Big(|\bar{D}_{1}^{0}D^{*0}\rangle-c|\bar{D}^{*0}D_{1}^0\rangle\Big)\\
&&+\Big(|{D}_{s1}^-
D_s^{*+}\rangle-c|D_s^{*-}D_{s1}^+\rangle\Big)\Big].
\end{eqnarray*}

We use $Z^0$ in the $J=0$ case as an example to illustrate how to
determine $c$. At the quark level, this molecular state may be
written as
\begin{equation}
J_{Z^0}=\frac{1}{2}[J_1-cJ_2-(J_3-cJ_4)]
\end{equation}
where
\begin{eqnarray*}
J_1=(\bar{c}^a\gamma_{\mu}\gamma_5 d^a)(\bar{d}^e\gamma^{\mu}c^e),\\
J_{2}=(\bar{d}^a\gamma_{\mu}\gamma_5 c^a)(\bar{c}^e\gamma^{\mu}d^e),\\
J_3=(\bar{c}^a\gamma_{\mu}\gamma_5 u^a)(\bar{u}^e\gamma^{\mu}c^e),\\
J_{4}=(\bar{u}^a\gamma_{\mu}\gamma_5
c^a)(\bar{c}^e\gamma^{\mu}u^e)\;.
\end{eqnarray*}
In the above equation, a and e are the color indices. Under charge
conjugate transformation, we have
\begin{eqnarray*}
{\hat C}J_1 {\hat C}^{-1}
=-(\bar{d}^a\gamma_{\mu}\gamma_5 c^a)(\bar{c}^e\gamma^{\mu}d^e)=-J_2,\\
{\hat C}J_2{\hat C}^{-1}=-(\bar{c}^a\gamma_{\mu}\gamma_5 d^a)
(\bar{d}^e\gamma^{\mu}c^e)=-J_1,\\
{\hat C}J_3{\hat C}^{-1}=
-(\bar{u}^a\gamma_{\mu}\gamma_5 c^a)(\bar{c}^e\gamma^{\mu}u^e)=-J_4,\\
{\hat C}J_4{\hat C}^{-1}=-(\bar{c}^a\gamma_{\mu}\gamma_5
u^a)(\bar{u}^e\gamma^{\mu}c^e)=-J_3\; .
\end{eqnarray*}
Therefore, we get
\begin{equation}
{\hat C}J_{Z^0}{\hat C}^{-1}=\frac{1}{2}[-J_2+cJ_1-(-J_4+cJ_3)].
\end{equation}
In other words, the $C$-parity of $Z^0$ is $C=\pm$ for $c=\pm 1$.
Since $Z^+(4430)$ is a state with isospin 1 and $G = +$, one
requires $c = -1$ for these states. The quantum numbers of $Z^0$ are
$I^G(J^{PC})=1^+(0,1,2)^{--}$.

We want to emphasize that the presence of both the charm and
anti-charm quark (and the light quark and anti-quark) in the
expression of $J_i$ ensures there is no arbitrary phase factor
under charge conjugate transformation.

There is one intuitive and natural way to interpret the above
results if we consider the flavor SU(4) symmetry, which is of course
broken badly in reality. If we naively assume the flavor SU(4)
symmetry, then states within the same multiplet should carry the
same coefficient under charge conjugate transformation. For example,
${\hat C} |D^{\ast -}\rangle = C(D^{\ast -}) |D^{\ast +}\rangle$
where the coefficient $C(D^{\ast -})$ takes the same value as either
$C(\rho^0)$ or $C(J/\psi)$. I.e., $C(D^{\ast -})=-1$. Similarly,
$C(D_1^-)=+1$. In other words, the $c=-1$ case leads to the $Z^0$
state with negative $C$-parity.

In addition, one obtains the flavor wave functions of these states
with opposite $G$-parity if we take $c=+$ in the above equations.
For example, we will also discuss whether $\widetilde{Z}^{+}$ could
be a molecular state:
\begin{eqnarray*}
|\widetilde{Z}^{+}\rangle&=&\frac{1}{\sqrt{2}}\Big(|\bar{D}_{1}^0
D^{*+}\rangle-|\bar{D}^{*0}D_{1}^+\rangle\Big).
\end{eqnarray*}
If $\tilde{Z}^+$ exists, this state may be discovered in either
$J/\psi \pi^+\pi^0$ or $\psi'\pi^+\pi^0$ channel.

If we replace one of the c (or $\bar c$) by $b$ (or $\bar b$), we
can get molecular states such as $(b\bar q)-(\bar c q)$. With the
dual replacement $c\to b, \bar c\to \bar b$, we get the hidden
bottom molecular states $(b\bar q)-(\bar b q)$.

\section{Effective Lagrangians and coupling constants}\label{sec3}

We collect the effective chiral Lagrangian used in the derivation
of the OPEP in this section. In the chiral and heavy quark dual
limits, the Lagrangian relevant to our calculation reads
\cite{falk,casalbuoni}
\begin{eqnarray}
\mathcal{L}&=&ig {\rm Tr}[H_b {A}\!\!\!\slash_{ba}\gamma_5\bar{H}_a
]+ig'{\rm Tr}[ S_b {A}\!\!\!\slash_{ba}\gamma_5\bar{S}_a]
 \nonumber\\&&+ig''{\rm Tr}[T_{\mu b}
A\!\!\!\slash_{ba}\gamma_5\bar{T}_a^{\mu}]\nonumber\\
&&+[ih {\rm Tr}[S_b{A}\!\!\!\slash_{ba}\gamma_5
\bar{H}_a]+h.c.]\nonumber\\&&+\{i\frac{h_1}{\Lambda_{\chi}}{\rm
Tr}[T_b^{\mu}(D_{\mu}{A}\!\!\!\slash)_{ba}\gamma_5\bar{H}_a]+h.c.\}
\nonumber\\&& +\{i\frac{h_2}{\Lambda_{\chi}}{\rm
Tr}[T_b^{\mu}(D\!\!\!\!/A_{\mu})_{ba}\gamma_5\bar{H}_a]+h.c.\},\label{aa}
\end{eqnarray}
where
\begin{eqnarray}
H_a&=&\frac{1+\not v}{2 }[P_{a}^{*\mu}-P_a \gamma_5],\\
S_{a}&=&\frac{1+\not v}{2 }[P_{1a}^{'\mu}\gamma_{\mu}\gamma_5
-P_{0a}^{*}],\\
T_{a}^{\mu}&=&\frac{1+\not v}{2 }\Big\{P^{*\mu\nu}_{2a}
\gamma_{\nu}-\sqrt{\frac{3}{2}}P_{1a}^{\nu}\gamma_5 [g_{\nu}^{\mu}
\nonumber\\&&-\frac{1}{3}\gamma_{\nu}(\gamma^{\mu}-v^{\mu})]\Big\}
\end{eqnarray}
and the axial vector field $A_{ab}^{\mu}$ is defined as
\begin{eqnarray*}
A_{ab}^{\mu}=\frac{1}{2}(\xi^{\dag}\partial^{\mu}\xi-\xi\partial^{\mu}\xi^{\dag})_{ab}=
\frac{i}{f_{\pi}}\partial^{\mu}\mathcal{M}+\cdots
\end{eqnarray*}
with $\xi=\exp(i\mathcal{M}/f_{\pi})$, $f_\pi=132$ MeV and
\begin{eqnarray}
\mathcal{M}&=&\left(\begin{array}{ccc}
\frac{\pi^{0}}{\sqrt{2}}+\frac{\eta}{\sqrt{6}}&\pi^{+}&K^{+}\\
\pi^{-}&-\frac{\pi^{0}}{\sqrt{2}}+\frac{\eta}{\sqrt{6}}&
K^{0}\\
K^- &\bar{K}^{0}&-\frac{2\eta}{\sqrt{6}}
\end{array}\right).
\end{eqnarray}

After expanding Eq. (\ref{aa}) to the leading order of the pion
field, we further obtain
\begin{eqnarray}
\mathcal{L}_{D^{*+}D^{*+}\pi^0}&=&g_{D^{*+}D^{*+}\pi^0}\epsilon^{\alpha\beta\mu\nu}
{D^{*+}_{\alpha}}(\partial_{\mu}\pi^0)(\partial_{\nu}{D}^{*-}_{\beta})+h.c.,\nonumber\\\label{la-1}\\
\mathcal{L}_{D_1^{'0}D_1^{'0}\pi^0}&=&g_{D_1^{'0}D_1^{'0}\pi^0}\epsilon^{\alpha\beta\mu\nu}D^{'0}_{1\alpha}
(\partial_{\mu}\pi^0)(\partial_{\nu} \bar{D}^{'0}_{1\beta})+h.c.,\nonumber\\\label{la-2}\\
\mathcal{L}_{D_1^{0}D_1^0\pi^0}&=&g_{D_1^{0}D_1^0\pi^0}\epsilon^{\alpha\beta\mu\nu}
({D}_{1\alpha}^0)(\partial_{\mu}\pi^0)(\partial_{\nu}{\bar{D}_{1\beta}^0})+h.c.,\nonumber\\\label{la-3}\\
\mathcal{L}_{D^{*}D_1^{'}\pi^0}&=&g_{D^{*}D_1^{'}\pi^0}\big[-(\partial^{\alpha}D^{*}_{\beta})(\partial^{\beta}\pi^0)
D_{1\alpha}^{'}\nonumber\\&&+D_{\beta}^{*}(\partial^{\alpha}\pi^0)(\partial^{\beta}D_{1\alpha}^{'})+(\partial^{\mu}
D_{\alpha}^{*})(\partial_{\mu}\pi^0)D_{1}^{'\alpha}\big]\nonumber\\&&+h.c.
\label{la-4}
\end{eqnarray}
\begin{eqnarray}
\mathcal{L}_{D*D_1\pi^0}=&&g_{D^*D_1\pi^0}\big[D^{*}_{\beta}D_{1\nu}
g^{\lambda\nu}(\partial^{\beta}\partial_{\lambda}\pi^0)\nonumber\\
&&-D^{*}_{\beta}D_{1\nu}g^{\beta\nu}(\partial^{\lambda}\partial_{\lambda}\pi^0)
 +2D^{*}_{\beta}D_{1\nu}g^{\beta\lambda}(\partial^{\nu}\partial_{\lambda}\pi^0)
\nonumber\\&&+\frac{1}{m_{D^*}m_{D_1}}(\partial^{\lambda}D^{*\nu})(\partial_{\alpha}\partial_{\lambda}\pi^0)(\partial^{\alpha}D_{1\nu})\big],\label{la-5}
\end{eqnarray}
where $D_{1}'$ denotes the P-wave axial-vector state in the
$(0^+,1^+)$ doublet while $D_{1}$ is the $1^+$ state in the
$(1^+,2^+)$ doublet. The coupling constants
$g_{D^{*+}D^{*+}\pi^0}$, $g_{D_1^{'0}D_1^{'0}\pi^0}$,
$g_{D_1^{0}D_1^0\pi^0}$ and $g_{D^*D_1^{(')}\pi^0}$ are
\begin{eqnarray}
&&g_{D^{*+}D^{*+}\pi^0}=-\frac{\sqrt{2}g}{f_{\pi}},\;\;\;g_{D_1^{'0}D_1^{'0}\pi^0}=\frac{\sqrt{2}g'}{f_{\pi}},
\nonumber\\&&
g_{D_1^{0}D_1^0\pi^0}=-\frac{5g''}{3\sqrt{2}f_{\pi}},
\nonumber\\&&g_{D^{*+}D_1^{'+}\pi^0}=g_{D^{*0}D_1^{'0}\pi^0}=-\frac{i\sqrt{2}h}{f_{\pi}},\nonumber\\&&
g_{D^{*0}D_{1}^{0}\pi^0}=-g_{D^{*+}D_{1}^{+}\pi^0}=-\frac{\sqrt{m_{D^*}m_{D_1}}}{\sqrt{3}f_{\pi}\Lambda_{\chi}}(h_1+h_2).\nonumber
\end{eqnarray}

The coupling constant $g$ was studied in many theoretical
approaches such as QCD sum rules \cite{QSR,QSR-1,QSR-2,QSR-3} and
quark model \cite{falk}. In this work, we use the value $g=0.59\pm
0.07\pm0.01$ extracted by fitting the experimental width of $D^*$
\cite{isoda}. Falk and Luke obtained an approximate relation
$|g'|=|g|/3$ and $|g''|=|g|$ in quark model \cite{falk}. However
the phase between $g'$ and $g''$ is not fixed. With the available
experimental information, Casalbuoni and collaborators extracted
$h=-0.56\pm 0.28$ and $h'=(h_1+h_2)/\Lambda_{\chi}=0.55$
GeV$^{-1}$ \cite{casalbuoni}. If we replace the meson field in the
heavy quark limit in the above equations by the fields in full QCD
and scale the coupling constants by a factor $\sqrt{m_{D^\ast}}$ etc,
we get the effective Lagrangian in full QCD, which is used below in
the derivation of the potential.

\section{Derivation of the one pion exchange potential\label{sec4}}

Study of the possible molecular states, especially the system of a
pair of heavy mesons, started more than three decades ago. The
presence of the heavy quarks lowers the kinetic energy while the
interaction between two light quarks could still provide strong
attraction. Okun and Voloshin proposed possibilities of
the molecular states involving charmed mesons \cite{Okun}. Rujula,
Geogi and Glashow once suggested $\psi(4040)$ as a $D^*\bar{D}^*$
molecular state \cite{RGG}. T\"{o}rnqvist studied possible
deuteronlike two-meson bound states such as $D\bar{D}^*$ and
$D^*\bar{D}^*$ using a quark-pion interaction model \cite{Tornqvist}.
Dubynskiy and Voloshin proposed
that there exists a possible new resonance at the $D^*\bar{D}^*$
threshold \cite{voloshin-1,voloshin}.

Several groups suggested $X(3872)$ could be a good molecular
candidate \cite{close-m,voloshin-m,wong-m,swanson,torn-m}.
However, Suzuki argued that $X(3872)$ is not a molecule state of
$D^0\bar{D}^{*0}+\bar{D}^0 D^{*0}$ \cite{suzuki}. Instead,
$X(3872)$ may have a dominant $c\bar{c}$ component with some
admixture of $D^0\bar{D}^{*0}+\bar{D}^0 D^{*0}$
\cite{chao-3872,suzuki,zhu-review}.

\begin{figure}[htb]
\begin{tabular}{cccccccc}
\scalebox{0.5}{\includegraphics{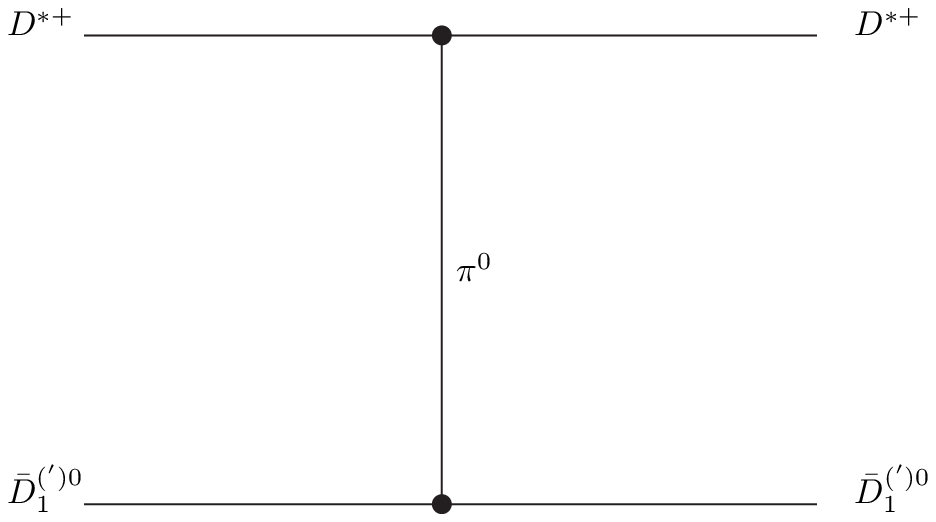}}\\\\(a)\\\\
\scalebox{0.5}{\includegraphics{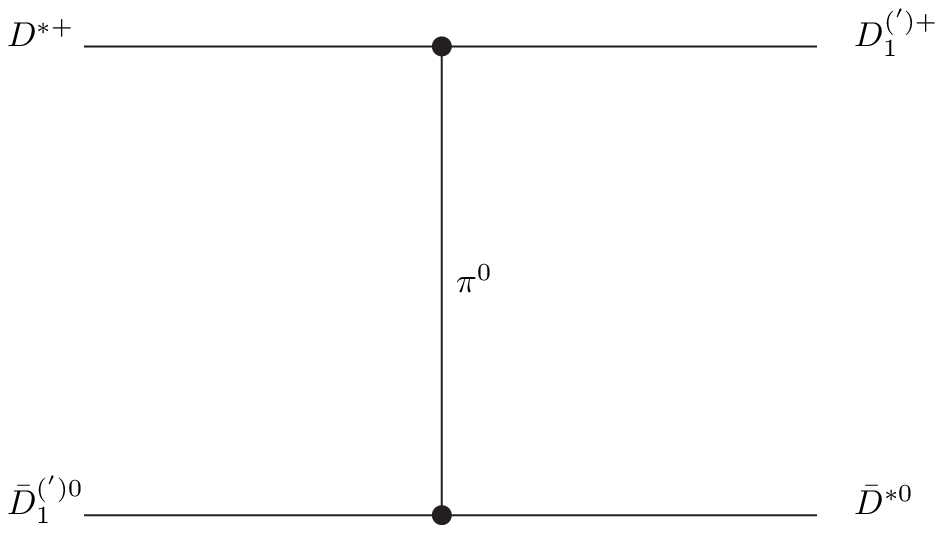}}\\\\(b)
\end{tabular}\caption{(a) The
single pion exchange in the direct scattering process
$D^{*+}\bar{D}_{1}^{(')0}\to D^{*+}\bar{D}_{1}^{(')0}$. (b) The
crossed process $D^{*+}\bar{D}_{1}^{(')0}\to
D_1^{(')+}\bar{D}^{*0}$. \label{figure}}
\end{figure}

In this work we will explore whether $Z^+(4430)$ could be a S-wave
molecular state of $D^*$ and $\bar{D}_1'$ (or $\bar{D}_1$), which
is loosely bound by the long-range pion exchange potential. We
first derive the scattering matrix elements between the pair of
$D$ mesons as shown in Fig. \ref{figure}. With the Breit
approximation, we can get the one-pion exchange potentials. Since
the flavor wave function of $Z^+$ contains two components, we have
to consider both the direct scattering diagram Fig. \ref{figure}
(a) and the crossed diagram Fig. \ref{figure} (b). Note only the
crossed diagram contributes to OPEP in the case of $X(3872)$.

Recall the flavor wave function of $Z^{+}(4430)$ reads
\begin{eqnarray}
|Z^{+}\rangle=\frac{1}{\sqrt{2}}\big[|A'\rangle+|B'\rangle\big]
\end{eqnarray}
with $|A'\rangle=|\bar{D}_{1}'^{0}D^{*+}\rangle$ and
$|B'\rangle=|D_{1}'^{+}\bar{D}^{*0}\rangle$ or
\begin{eqnarray}
|Z^{+}\rangle=\frac{1}{\sqrt{2}}\big[|A\rangle+|B\rangle\big]
\end{eqnarray}
with $|A\rangle=|\bar{D}_{1}^{0}D^{*+}\rangle$ and
$|B\rangle=|D_{1}^{+}\bar{D}^{*0}\rangle$. The mass of $Z^+(4430)$
is expressed as
\begin{eqnarray}\label{energy}
M_{Z^+}=m_{D^*}+m_{D_1^{(')}}+ T +E+\delta,
\end{eqnarray}
where $T$ is the kinetic energy in the center of mass frame,
$E=\langle D^{*+}\bar{D}_{1}^{(')0} |\mathcal{H}_1|
D^{*+}\bar{D}_{1}^{(')0}\rangle$ and $\delta=\langle
D^{*+}\bar{D}_{1}^{(')0}|\mathcal{H}_2|D_1^{{'}+}\bar{D}^{*0}\rangle$.
$\mathcal{H}_1$ and $\mathcal{H}_2$ correspond to the interaction in
Fig. \ref{figure} (a) and (b) respectively. For the possible
${\tilde Z}^+$ state with negative G-parity, we can get its mass
through the replacement $+\delta \to -\delta$ in Eq. (\ref{energy}).

\subsection{Scattering amplitudes}

We collect the scattering amplitudes in the different channels
below. For the process
$D^{*+}(p_1,\epsilon_1)\bar{D}_{1}^{'0}(p_2,\epsilon_2)\to
D^{*+}(p_3,\epsilon_3)\bar{D}_{1}^{'0}(p_4,\epsilon_4)$, the
amplitude is
\begin{eqnarray}
&&i\mathcal{M}[D^{*+}(p_1,\epsilon_1)\bar{D}_{1}^{'0}(p_2,\epsilon_2)\to
D^{*+}(p_3,\epsilon_3)\bar{D}_{1}^{'0}(p_4,\epsilon_4)]\nonumber\\&&=
\frac{2igg'}{f_{\pi}^2}\frac{1}{q^2-m_{\pi}^2}\varepsilon^{\alpha\beta\mu\nu}\varepsilon^{\alpha'\beta'\mu'\nu'}
q_{\mu}q_{\mu'}p_{1\nu}p_{1\nu'}\nonumber\\&&\times(\epsilon^{\lambda_1}_{1\alpha}\epsilon^{\lambda_2}_{2\alpha'})
(\epsilon^{\lambda_1'}_{3\beta}\epsilon^{\lambda_2'}_{4\beta'}).
\end{eqnarray}
For $D^{*+}(p_1,\epsilon_1)\bar{D}_{1}^{0}(p_2,\epsilon_2)\to
D^{*+}(p_3,\epsilon_3)\bar{D}_{1}^{0}(p_4,\epsilon_4)$, one gets
\begin{eqnarray}
&&i\mathcal{M}[D^{*+}(p_1,\epsilon_1)\bar{D}_{1}^{0}(p_2,\epsilon_2)\to
D^{*+}(p_3,\epsilon_3)\bar{D}_{1}^{0}(p_4,\epsilon_4)]\nonumber\\&&=
-\frac{5igg''}{3f_{\pi}^2}\frac{1}{q^2-m_{\pi}^2}\varepsilon^{\alpha\beta\mu\nu}\varepsilon^{\alpha'\beta'\mu'\nu'}
q_{\mu}q_{\mu'}p_{1\nu}p_{1\nu'}\nonumber\\&&\times(\epsilon^{\lambda_1}_{1\alpha}\epsilon^{\lambda_2}_{2\alpha'})
(\epsilon^{\lambda_1'}_{3\beta}\epsilon^{\lambda_2'}_{4\beta'}).
\end{eqnarray}
For $D^{*+}(p_1,\epsilon_1)\bar{D}_{1}^{'0}(p_2,\epsilon_2)\to
{D}_{1}^{'+}(p_3,\epsilon_3)\bar{D}^{*0}(p_4,\epsilon_4)$, the
amplitude is
\begin{eqnarray}
&&i\mathcal{M}[D^{*+}(p_1,\epsilon_1)\bar{D}_{1}^{'0}(p_2,\epsilon_2)\to
{D}_{1}^{'+}(p_3,\epsilon_3)\bar{D}^{*0}(p_4,\epsilon_4)]\nonumber\\&&=
\frac{2ih^2}{f_{\pi}^2}\frac{1}{q^2-m_{\pi}^2}[-p_1^{\alpha}q^{\beta}-q^{\alpha}p_3^{\beta}+(p_1\cdot
q)g^{\alpha\beta}]\nonumber\\&&\times[-p_4^{\alpha'}q^{\beta'}-q^{\alpha'}p_{2}^{\beta'}+(p_4\cdot
q)g^{\alpha'\beta'}](\epsilon^{\lambda_1}_{1\beta}\epsilon^{\lambda_2}_{2\alpha'})
(\epsilon^{\lambda_1'}_{3\alpha}\epsilon^{\lambda_2'}_{4\beta'}).\nonumber\\
\end{eqnarray}
The amplitude of the process
$D^{*+}(p_1,\epsilon_1)\bar{D}_{1}^{0}(p_2,\epsilon_2)\to
{D}_{1}^{+}(p_3,\epsilon_3)\bar{D}^{*0}(p_4,\epsilon_4)$ is
\begin{eqnarray}
&&i\mathcal{M}[D^{*+}(p_1,\epsilon_1)\bar{D}_{1}^{0}(p_2,\epsilon_2)\to
{D}_{1}^{+}(p_3,\epsilon_3)\bar{D}^{*0}(p_4,\epsilon_4)]\nonumber\\&&=
\frac{-im_{D^*}m_{D_1}(h_1+h_2)^2}{3f_{\pi}^2
\Lambda^2_{\chi}}\frac{1}{q^2-m_{\pi}^2}\Big[3q^{\nu}q^{\beta}-g^{\beta\nu}q^2\nonumber\\&&+\frac{g^{\beta\nu}}{m_{D^*}m_{D_1}}(p_1\cdot
q)(q\cdot
p_3)\Big]\Big[3q^{\nu'}q^{\beta'}-g^{\beta'\nu'}q^2\nonumber\\&&
+\frac{g^{\beta'\nu'}}{m_{D^*}m_{D_1}}(p_2\cdot q)(q\cdot
p_4)\Big](\epsilon^{\lambda_1}_{1\beta}\epsilon^{\lambda_2}_{2\nu'})
(\epsilon^{\lambda_1'}_{3\nu}\epsilon^{\lambda_2'}_{4\beta'}).\nonumber\\
\end{eqnarray}
Here the polarization vector is defined as
$\epsilon^{\pm1}=\frac{1}{\sqrt{2}}(0,\pm1,i,0)$ and
$\epsilon^{0}=(0,0,0,-1)$.

\subsection{The one-pion-exchange potential}

We impose the constraint on the scattering amplitudes that initial
states and final states should have the same angular momentum. The
molecular state $|J,J_z\rangle$ composed of the $1^-$ and $1^+$
charm meson pair can be constructed as
\begin{equation}
|J,J_z\rangle=\sum_{\lambda_1,\lambda_2}\langle
1,\lambda_1;1,\lambda_2|J,J_z \rangle|p_1,\epsilon_1;
p_2,\epsilon_2\rangle
\end{equation}
where $\langle1,\lambda_1;1,\lambda_2|J,J_z\rangle$ is the
Clebsch-Gordan coefficient. Combining the equation with the
scattering amplitudes, one gets the matrix element
$i\mathcal{M}(J,Jz)$.

With the Breit approximation, the interaction potential in the
momentum space is related to $i\mathcal{M}(J,Jz)$
\begin{equation}
V(q)=-\frac{1}{\sqrt{\prod_{i}2m_i \prod_{f}2m_f }}\mathcal{M}(J,Jz)
\end{equation}
where $m_i$ and $m_f$ denote the masses of the initial and final
states respectively. We collect the expressions of the potential
in Tables \ref{p-space1} and \ref{p-space2}. We have explicitly
shown the resulting potentials are the same for the different
$J_z$ component.

\begin{center}
\begin{table}[htb]
\caption{The one-pion-exchange potential between $D^*$
($\bar{D}^*$) and $\bar{D}_1^{'}$ ($D_1^{'}$). Here the
expressions are for the $J_z=0$ components.
$A'=\bar{D}_{1}'^{0}D^{*+}$ and $B'=D_{1}'^{+}\bar{D}^{*0}$.
\label{p-space1}}
\begin{tabular}{c||c|c}
\hline State&$A^\prime(B^\prime)\rightarrow A^\prime(B^\prime)
$&$A^\prime (B^\prime)\rightarrow B^\prime(A^\prime)$\\
\hline\hline
$0^-$&$ \frac{gg'}{3 f_\pi^2}\frac{{\mathbf{q}}^2}{{\mathbf{q}}^2+m_\pi^2} $&$-\frac{h^2}{2f_\pi^2}\frac{(q^0)^2}{(q^0)^2-{\mathbf{q}}^2-m_\pi^2}$\\
\hline
$1^-$&$\frac{gg'}{2 f_\pi^2}\frac{q_z^2}{{\mathbf{q}}^2+m_\pi^2} $&$-\frac{h^2}{2f_\pi^2}\frac{(q^0)^2}{(q^0)^2-{\mathbf{q}}^2-m_\pi^2}$\\
\hline
$2^-$&$-\frac{gg'}{6 f_\pi^2}\frac{2{\mathbf{q}}^2-3q_z^2}{{\mathbf{q}}^2+m_\pi^2}  $&$-\frac{h^2}{2f_\pi^2}\frac{(q^0)^2}{(q^0)^2-{\mathbf{q}}^2-m_\pi^2}$\\
\hline
\end{tabular}
\end{table}\end{center}

\begin{center}
\begin{table}[htb]
\caption{The one-pion-exchange potential between $D^*$
($\bar{D}^*$) and $\bar{D}_1$ ($D_1$). In this table,
$A=\bar{D}_{1}^{0}D^{*+}$ and $B=D_1^{+}\bar{D}^{*0}$.
\label{p-space2}}
\begin{tabular}{c||c|c}
\hline State&$A(B)\rightarrow A(B)
$&$A (B)\rightarrow B(A)$\\
\hline\hline
$0^-$&$ -\frac{5gg''}{18 f_\pi^2}\frac{{\mathbf{q}}^2}{{\mathbf{q}}^2+m_\pi^2} $&$\frac{h'^2}{6 f_\pi^2} \frac{({\mathbf{q}}^2)^2}{(q^0)^2-{\mathbf{q}}^2-m_\pi^2}$\\
\hline
$1^-$&$-\frac{5gg''}{12 f_\pi^2}\frac{q_z^2}{{\mathbf{q}}^2+m_\pi^2} $&$-\frac{h'^2}{12 f_\pi^2} \frac{2({\mathbf{q}}^2)^2-3q_z^2{\mathbf{q}}^2}{(q^0)^2-{\mathbf{q}}^2-m_\pi^2}$\\
\hline
$2^-$&$\frac{5gg''}{36 f_\pi^2}\frac{2{\mathbf{q}}^2-3q_z^2}{{\mathbf{q}}^2+m_\pi^2}  $&$\frac{h'^2}{8 f_\pi^2}\frac{({\mathbf{q}}^2)^2-8q_z^2{\mathbf{q}}^2+9q_z^4}{(q^0)^2-{\mathbf{q}}^2-m_\pi^2}$\\
\hline
\end{tabular}
\end{table}\end{center}

\begin{center}\begin{table}[htb]\caption{The one pion exchange potential
in the coordinate space with $A'=\bar{D}_{1}'^{0}D^{*+}$ and
$B'=D_{1}'^{+}\bar{D}^{*0}$. \label{t1}}
\begin{tabular}{c||c|cccc}
\hline State&$A^\prime(B^\prime)\rightarrow A^\prime(B^\prime)
$&{$A^\prime (B^\prime)\rightarrow B^\prime(A^\prime)$}\\
\hline\hline
$0^-$&$ \frac{gg'}{3 f_\pi^2}[\delta(\mathbf{r})-\frac{m_\pi^2}{4\pi r}e^{-m_\pi r}] $&$\frac{h^2 (q^0)^2}{8\pi f_\pi^2}\frac{\cos(\mu r)}{r}$\\
\hline $1^-$&$\frac{gg'}{6
f_\pi^2}[\delta(\mathbf{r})-\frac{m_\pi^2}{4\pi r}e^{-m_\pi r}]
$&$\frac{h^2 (q^0)^2}{8\pi f_\pi^2}\frac{\cos(\mu r)}{r}$\\ \hline
$2^-$&$-\frac{gg'}{6
f_\pi^2}[\delta(\mathbf{r})-\frac{m_\pi^2}{4\pi r}e^{-m_\pi r}]
$&$\frac{h^2 (q^0)^2}{8\pi f_\pi^2}\frac{\cos(\mu r)}{r}$\\ \hline
\end{tabular}
\end{table}\end{center}

\subsubsection{OPEP from the direct scattering diagram}

From Tables \ref{p-space1} and \ref{p-space2}, it's very
interesting to note that the OPEP from the direct scattering
diagram in the $J^P=0^-$ and $1^-$ channel always has the same
sign while it's opposite to that in the $J^P=2^-$ channel. In
other words, there must be attraction in one of three channels no
matter what sign $gg'$ takes.

After making the Fourier transformation, we get the
one-pion-exchange potential in the configuration space. The final
potentials are obtained with the following replacements:
$x^2\rightarrow r^2/3$, $x^4\rightarrow r^4/5$, $x^2y^2\rightarrow
r^4/15$ etc since we consider only the S-wave system.
Alternatively, one may average the potential in the momentum space
first.

In the derivation of the OPEP, we let $q^0\approx 0$ as usually
done for the direct scattering diagram in Fig. \ref{figure} (a).
The resulting potential from the direct scattering diagram is the
familiar Yukawa potential plus a $\delta$ function, similar to
OPEP in the nucleon-nucleon potential.

\begin{widetext}
\begin{center}\begin{table}[htb]\caption{The one pion exchange potential
in the coordinate space with $A=\bar{D}_{1}^{0}D^{*+}$ and
$B=D_{1}^{+}\bar{D}^{*0}$.\label{t2}}
\begin{tabular}{c||c|c}
\hline State&$A(B)\rightarrow A(B)
$&$A (B)\rightarrow B(A)$\\
\hline\hline $0^-$&$ -\frac{5gg''}{18
f_\pi^2}[\delta(\mathbf{r})-\frac{m_\pi^2}{4\pi r}e^{-m_\pi r}]
$&$\frac{h'^2}{6 f_\pi^2}[\nabla^2\delta(\mathbf{r})
-\mu^2\delta(\mathbf{r})-\frac{\mu^4}{4\pi}\frac{\cos \mu r}{r}]$\\
\hline $1^-$&$-\frac{5gg''}{36
f_\pi^2}[\delta(\mathbf{r})-\frac{m_\pi^2}{4\pi r}e^{-m_\pi r}]
$&$-\frac{h'^2}{12 f_\pi^2}[\nabla^2\delta(\mathbf{r})
-\mu^2\delta(\mathbf{r})-\frac{\mu^4}{4\pi}\frac{\cos \mu r}{r}]$\\
\hline $2^-$&$\frac{5gg''}{36
f_\pi^2}[\delta(\mathbf{r})-\frac{m_\pi^2}{4\pi r}e^{-m_\pi r}]
$&$\frac{h'^2}{60 f_\pi^2}[\nabla^2\delta(\mathbf{r})
-\mu^2\delta(\mathbf{r})-\frac{\mu^4}{4\pi}\frac{\cos \mu r}{r}]$\\
\hline
\end{tabular}
\end{table}
\end{center}
\end{widetext}

\begin{figure}[htb]
\begin{tabular}{cccccccc}
\scalebox{0.8}{\includegraphics{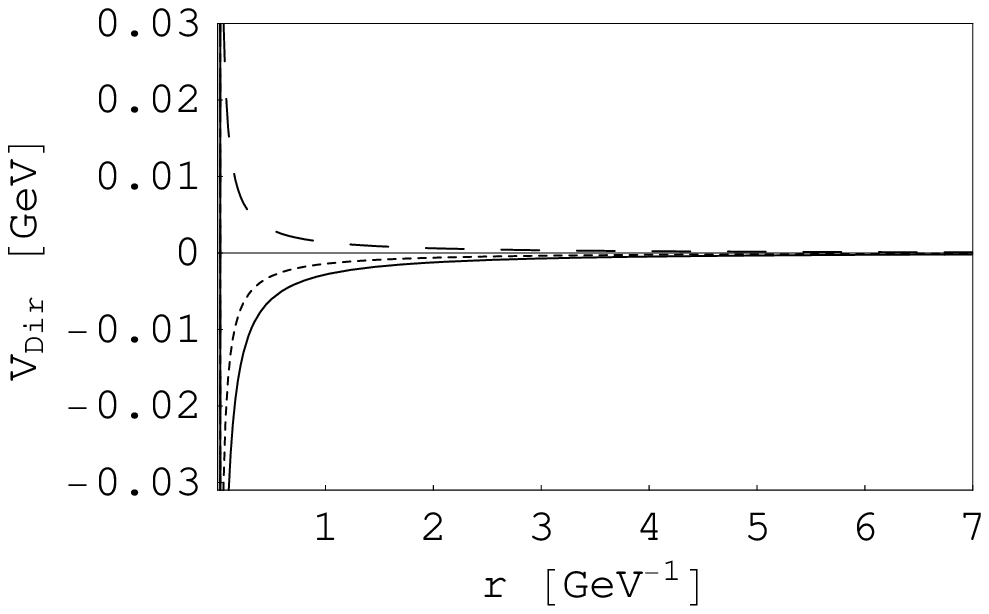}}\\\\(a)\\\\
\scalebox{0.8}{\includegraphics{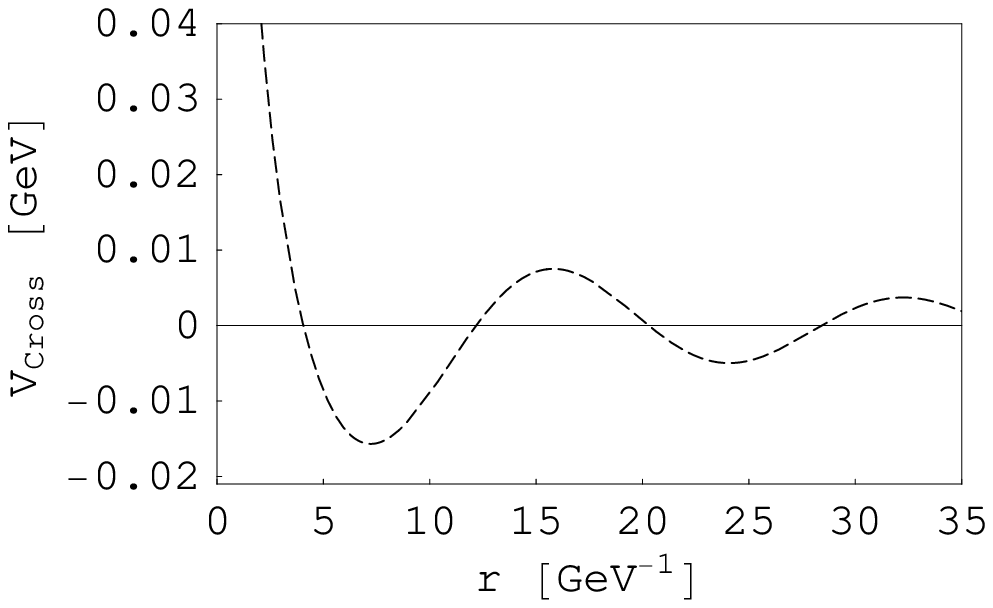}}\\\\(b)
\end{tabular}\caption{(a) The solid, dotted and dashed lines
correspond to the potentials from the direct scattering diagram
$A(B)\to A(B)$ in the $J^P=0^-, 1^-, 2^-$ channels respectively,
where $g=0.59$, $g'=g/3$ and $g''=g$. (b) The potential from the
crossed diagram $A(B)\to B(A)$.\label{poten-1} }
\end{figure}

\begin{figure}[htb]
\begin{tabular}{cccccccc}
\scalebox{0.8}{\includegraphics{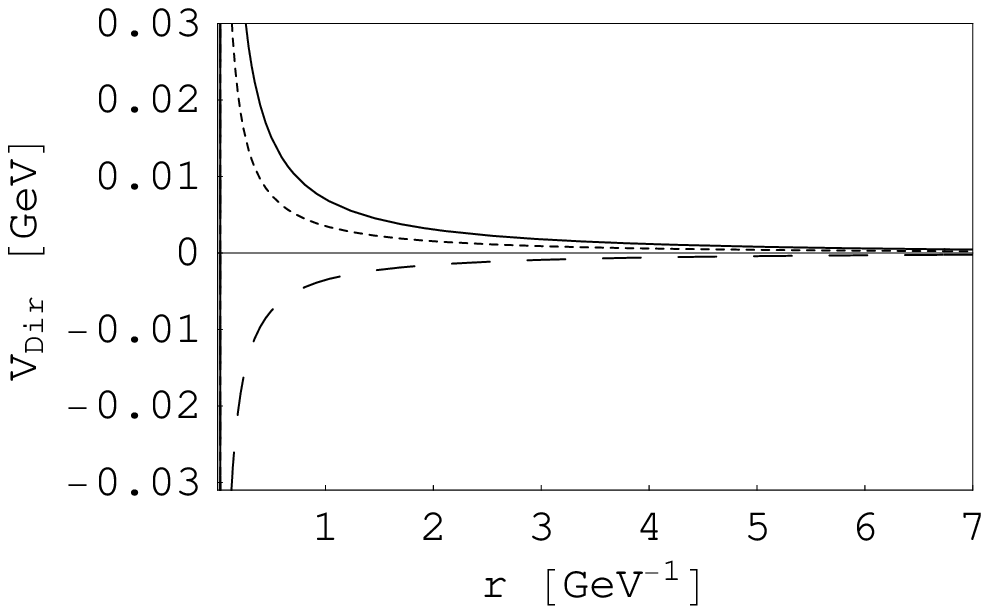}}\\\\(a)\\\\
\scalebox{0.8}{\includegraphics{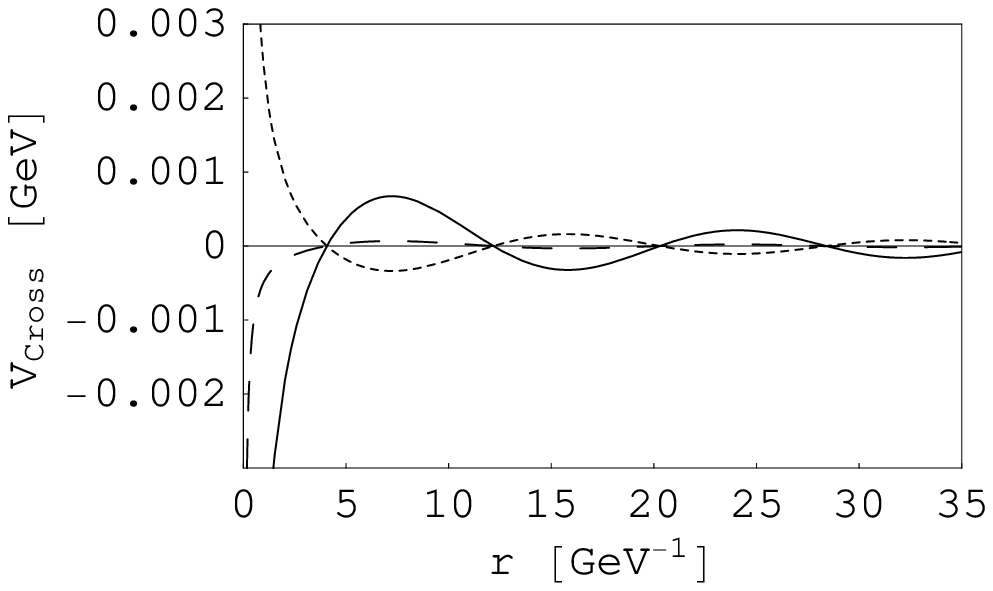}}\\\\(b)
\end{tabular}\caption{(a) The solid, dotted and dashed lines
correspond to the potentials from the direct scattering diagram
$A'(B')\to A(B')$ in the $J^P=0^-, 1^-, 2^-$ channels
respectively, where $g=0.59$, $g'=g/3$ and $g''=g$. (b) The
potential from the crossed diagram $A'(B')\to B'(A')$.
\label{poten-2}}
\end{figure}

\subsubsection{OPEP from the crossed diagram}

First we note from Table \ref{p-space2} that the OPEP from the
crossed diagram contains terms such as $\mathbf{q}^4$ in the
numerator, which reflects the fact that $D_1$ decays into
$D^\ast\pi$ via D-wave. But the approximation $q^0\approx 0$ is
not reasonable for the crossed diagram in Fig. \ref{figure} (b)
because $q^0\approx M_{D_1}-M_{D^\ast}\approx 410
\;\mbox{MeV}\approx 3 m_\pi$. In the following, we take $J^P=0^-$
as an example to illustrate how to deal with the crossed diagram.
Since the $D_1^\prime$ meson decays into $D^\ast \pi$, the crossed
diagram contains a small imaginary part, which is roughly of the
order $\Gamma_{D_1^\prime}$. However only the real part of the
scattering amplitude contributes to the potential. Hence the
principal integration is always assumed throughout the Fourier
transformation for the crossed diagram.
\begin{eqnarray*}
V(\mathbf{r})&=&\int {\cal {\hat P}}\Big [
-\frac{h^2}{2f_\pi^2}\frac{(q^0)^2}{(q^0)^2-{\mathbf{q}}^2-m_\pi^2+i\epsilon}
e^{i\mathbf{q}\cdot \mathbf{r}}\Big] \frac{d
\mathbf{q}}{(2\pi)^3}\\&=& -\frac{h^2}{2f_\pi^2}\int {\cal {\hat
P}}\Big [ \frac{(q^0)^2}{\mu^2-{\mathbf{q}}^2+i\epsilon }
e^{i\mathbf{q}\cdot \mathbf{r}} \Big] \frac{ d
\mathbf{q}}{(2\pi)^3}\\&=&\frac{h^2 (q^0)^2}{8\pi
f_\pi^2}\frac{\cos(\mu r)}{r}
\end{eqnarray*}
where $\mu^2=(q^0)^2-m_{\pi}^2$ and $q^0\approx
M_{D_1}-M_{D^\ast}$. The resulting potential is of very long range
and oscillating. We list the expressions of the potential in the
coordinate space in Tables \ref{t1} and \ref{t2}.

For the $\bar{D}^*D_1^{'}$ system, it's clear from Table \ref{t1}
that the OPEP from the crossed diagram has the same overall sign
in all three channels. Moreover, the sign is positive. Because of
the factor $q_0^2\sim 9 m_\pi^2$, the OPEP from the crossed
diagram is numerically much larger than that from the direct
scattering diagram. For the $\bar{D}^*D_1$ case, we note from
Table \ref{t2} that the overall sign in the $1^-$ channel is
different from that in both $0^-$ and $2^-$ channels.

With the coupling constants $g=0.59$, $g'=g/3$ and $g''=g$, the
variation of the potential listed in Tables \ref{t1} and \ref{t2}
with $r$ (in unit of GeV$^{-1}$) is shown in Figs. \ref{poten-1}
and \ref{poten-2}.

We also note that the OPEP does not depend on the mass of the
charmed mesons. In other words, the same
potential may be used in the discussion of the hidden bottom
molecular states $(b\bar q)-(\bar b q)$.

\section{Results and discussions \label{sec5}}

Besides those coupling constants in Section \ref{sec3}, we also
need the following parameters in our numerical analysis:
$m_{D^*}=2007$ MeV, $m_{D_1'}=2430$ MeV, $m_{D_1}=2420$ MeV,
$m_{B^{*}}=5325$ MeV, $m_{B_{1}'}=5732$ MeV, $f_{\pi}=132$ MeV,
$m_{\pi}=135$ MeV \cite{PDG}; $m_{B_1}=5725$ MeV \cite{PDG-1}.

With the potentials derived above, we use the variational method
to investigate whether there exists a loosely bound state.
Our criteria of the formation of a
possible loosely bound molecular state is (1) the radial wave
function extend to 1 fm or beyond and (2) the minimum energy of
the system is negative. Our
trial wave functions include (a) $\psi(r)=(1+\alpha r)e^{-\alpha
r}$; (b) $\psi(r)=(1+\alpha r^2) e^{-\beta r^2}$; (c) $\psi(r)=
r^2(1+\alpha r) e^{-\beta r}$.

Unfortunately a solution satisfy the above criteria does not exist
for the system of $D^{'}_1-D^*$ or $D_1-D^*$ in all $J^P=0^-, 1^-,
2^-$ channels with the realistic coupling constants $g=0.59$,
$g'=g/3$ and $g''=g$ deduced from the width of $D^\ast, D_1$ and
$D_1^\prime$. Such a solution also does not exist if we switch the
sign of $gg'$ or enlarge the absolute value of $gg'$ by a factor
3. The same conclusion holds for the system of $B^{'}-B^*$,
$B-B^*$ and $\widetilde{Z}^+$ with negative $G$-parity.

In short summary, we have performed a dynamical study of the
$Z^+(4430)$ signal to see whether it is a loosely bound molecular
state of $D_1-D^*$ or $D_1'-D^*$. We find that the interaction
from the one pion exchange potential alone is not strong enough to
bind the pair of charmed mesons with realistic coupling constants.
Other dynamics is necessary if $Z^+(4430)$ is further established
as a molecular state by the future experiments.

It's interesting to note that the one pion exchange potential alone
does not bind the deuteron in nuclear physics either.  In fact, the
strong attractive force in the intermediate range is introduced in
order to bind the deuteron, which is sometimes modeled by the sigma
meson exchange. One may wonder whether the similar mechanism plays a
role in the case of $Z^+(4430)$ and $X(3872)$. Further work along
this direction is in progress.

If the $Z^+(4430)$ is really a $J^P =1^-$ molecular state, the
quantum number of its neutral partner state $Z^0(4430)$ is
$J^{PC}=1^{--}$. Such a state can be searched for in the $e^+e^-$
annihilation processes. Babar and Belle collaborations have observed
several new charmonium (or charmonium-like) states with
$J^{PC}=1^{--}$ around this mass range including $Y(4260)$,
$Y(4320)$, and $Y(4664)$ with the initial state radiation (ISR)
technique, although these states do not appear as a peak in the R
distribution. We strongly urge Belle and Babar collaborations to
search for the $Z^0(4430)$ state in the $\pi^0 \psi^\prime$ channel
using the ISR technique. The absence of a signal will be an
indication that the $J^P$ of $Z^+(4430)$ is not $1^-$.

\vfill
\section*{Acknowledgments}

We would like to thank Professor K.T. Chao, Professor Z.Y. Zhang,
and Dr C. Meng for useful discussions. This project was supported by
the National Natural Science Foundation of China under Grants
10421503, 10625521, 10675008, 10705001, 10775146 and the China
Postdoctoral Science foundation (20060400376).


\vfill

\begin{thebibliography}{99}
\bibitem{Belle-4430}Belle Collaboration, K. Abe et al.,
arXiv:0708.1790 [hep-ex].

\bibitem{Barbar-BR3872}Barbar Collaboration, B. Aubert et al., Phys. Rev. D 71, 071103
(2005).

\bibitem{Barbar-BR3872-06}Barbar Collaboration, B. Aubert et al., Phys. Rev. D 73, 011101 (2006).


\bibitem{rosner}J. L. Rosner, arXiv:0708.3496 [hep-ph].
\bibitem{maiani}L. Maiania, A.D. Polosab and V. Riquerb,
arXiv:0708.3997 [hep-ph].
\bibitem{Gershtein}S.S. Gershtein, A.K. Likhoded and G.P. Pronko,
arXiv:0709.2058 [hep-ph].
\bibitem{cky}K. Cheunga, W. Y. Keung and T. C. Yuan, arXiv:0709.1312
[hep-ph].
\bibitem{Qiao}C.F. Qiao, arXiv:0709.4066 [hep-ph].
\bibitem{qsr-lee}S. H. Lee, A. Mihara, F.S. Navarra and M. Nielsen,
arXiv:0710.1029 [hep-ph].
\bibitem{Bugg}D. V. Bugg, arXiv:0709.1254 [hep-ph].
\bibitem{PDG}W. M. Yao et al., Particle Data Group, J. Phys. G {\bf 33}, 1
(2006).
\bibitem{Meng}C. Meng and K. T. Chao, arXiv:0708.4222 [hep-ph].


\bibitem{falk}A. F. Falk and M. Luke, Phys. Lett. {\bf B 292}, 119
(1992).

\bibitem{casalbuoni}R. Casalbuoni, A. Deandrea, N. Di Bartolomeo,
F. Feruglio, R. Gatto and G. Nardulli, Phys. Rept. {\bf 281}, 145
(1997).


\bibitem{QSR}V.M. Belyaev, V. M. Braun, A. Khodjamirian and R. Ruckl,
Phys. Rev. {\bf D 51}, 6177 (1995).

\bibitem{QSR-1}F. S. Navarra, Marina Nielsen
and M.E. Bracco, Phys. Rev. {\bf D 65}, 037502 (2002).

\bibitem{QSR-2}F. S. Navarra, M. Nielsen, M. E. Bracco, M. Chiapparini
and C.L. Schat, Phys. Lett. {\bf B 489}, 319 (2000).

\bibitem{QSR-3}Y. B. Dai and S. L. Zhu, Eur. Phys. J. {\bf C 6}, 307 (1999).

\bibitem{isoda}C. Isola, M. Ladisa, G. Nardulli and P. Santorelli,
Phys. Rev. {\bf D 68}, 114001 (2003).


\bibitem{Okun}M. B. Voloshin and L. B. Okun, JETP Lett. {\bf 23}, 333 (1976).

\bibitem{RGG}A. D. Rujula, H. Georgi and S. L. Glashow, Phys. Rev.
Lett. {\bf 38}, 317 (1977).

\bibitem{Tornqvist} N.A. T\"{o}rnqvist, Nuovo Cim. {\bf A 107}, 2471-2476
(1994); Z.Phys. {\bf C 61}, 525-537 (1994).

\bibitem{voloshin-1}M. B. Voloshin, arXiv:hep-ph/0602233.

\bibitem{voloshin}S. Dubynskiy and M. B. Voloshin, Mod. Phys. Lett.
{\bf A 21}, 2779 (2006).

\bibitem{close-m}F. E. Close, P. R. Page, Phys. Lett. B 578, 119 (2004).

\bibitem{voloshin-m}M. B. Voloshin, Phys. Lett. B 579, 316 (2004).

\bibitem{wong-m}C. Y. Wong, Phys. Rev. C 69, 055202 (2004).

\bibitem{swanson}E. S. Swanson, Phys. Lett. B 588, 189 (2004); ibid B 598, 197 (2004).

\bibitem{torn-m}N. A. Tornqvist, Phys. Lett. B 590, 209 (2004).

\bibitem{suzuki}M. Suzuki, Phys. Rev. {\bf D 72}, 114013 (2005).

\bibitem{zhu-review}S. L. Zhu, hep-ph/0703225, to appear in Int. J. Mod.
Phys. E; arXiv:0707.4586 [hep-ph],

\bibitem{chao-3872}C. Meng, Y. J. Gao and K. T. Chao, arXiv:
hep-ph/0506222.

\bibitem{PDG-1}CDF Collaboration,
¡°Mass and width measurement of orbitally excited ($L=1$) $B^{**0}$
mesons¡±, CDF Note 8945,
www-cdf.fnal.gov/physics/new/bottom/070726.blessed-bss/; CDF
Collaboration, Andreas Gessler, the EPS talk given on HEP 2007 in
Manchester, arXiv:0709.3148 [hep-ex].




\end{thebibliography}
\end{document}